\documentclass{PoS}
\usepackage{amsmath,amssymb,amsfonts,graphics,amscd,amsfonts,epsf,epsfig,color}
\usepackage{bm}
\title{How to make a quantum black hole with ultra-cold gases}

\ShortTitle{How to make a quantum black hole with ultra-cold gases}

\author{Ippei Danshita\\
        Yukawa Institute for Theoretical Physics, Kyoto University, Kyoto 606-8502, Japan\\
        E-mail: \email{danshita@yukawa.kyoto-u.ac.jp}}

\author{\speaker{Masanori Hanada}\\
        Yukawa Institute for Theoretical Physics, Kyoto University, Kyoto 606-8502, Japan\\
        Stanford Institute for Theoretical Physics, Stanford University, Stanford, CA 94305, USA\\
        The Hakubi Center for Advanced Research, Kyoto University, Kyoto 606-8501, Japan\\
        E-mail: \email{hanada@yukawa.kyoto-u.ac.jp}}
        
\author{Masaki Tezuka\\
        Department of Physics, Kyoto University, Kyoto 606-8502, Japan\\
        E-mail: \email{tezuka@scphys.kyoto-u.ac.jp}}
                
\abstract{
The realization of quantum field theories on an optical lattice is an important subject 
toward the quantum simulation. 
We argue that such efforts would lead to the experimental realizations of quantum black holes. 
The basic idea is to construct non-gravitational systems which are equivalent to the quantum gravitational systems via the holographic principle. 
Here the `equivalence' means that two theories cannot be distinguished even in principle.
Therefore, if the holographic principle is true, one can create actual quantum black holes
by engineering the non-gravitational systems on an optical lattice.
In this presentation, we consider the simplest example: the Sachdev-Ye-Kitaev (SYK) model. 
We design an experimental scheme for creating the SYK model with use of ultra-cold fermionic atoms
such as $^6$Li.
          }

\FullConference{
		
		{\rm Unpublished manuscript originally intended for the proceedings of the
34th Annual Symposium on Lattice Field Theory, based on
our article arXiv:1606.02454
(published as Prog. Theor. Exp. Phys. 2017, 083I01 after the
proceedings deadline)}
}

\begin{document}

\section{Introduction}
A successful marriage of quantum mechanics and general relativity --- {\it quantum gravity} --- 
has been a dream of high energy theorists over decades. 
In this presentation, we will try to convince you that the efforts in a part of the lattice community 
--- the realization of quantum field theories on an optical lattice ---,  
combined with an idea from superstring theory, 
can possibly lead to a totally unexpected research subject: 
the experimental study of quantum gravity. 

\begin{figure}[htbp]
\begin{center}
\rotatebox{0}{
\scalebox{0.8}{ 
\includegraphics[width=10cm]{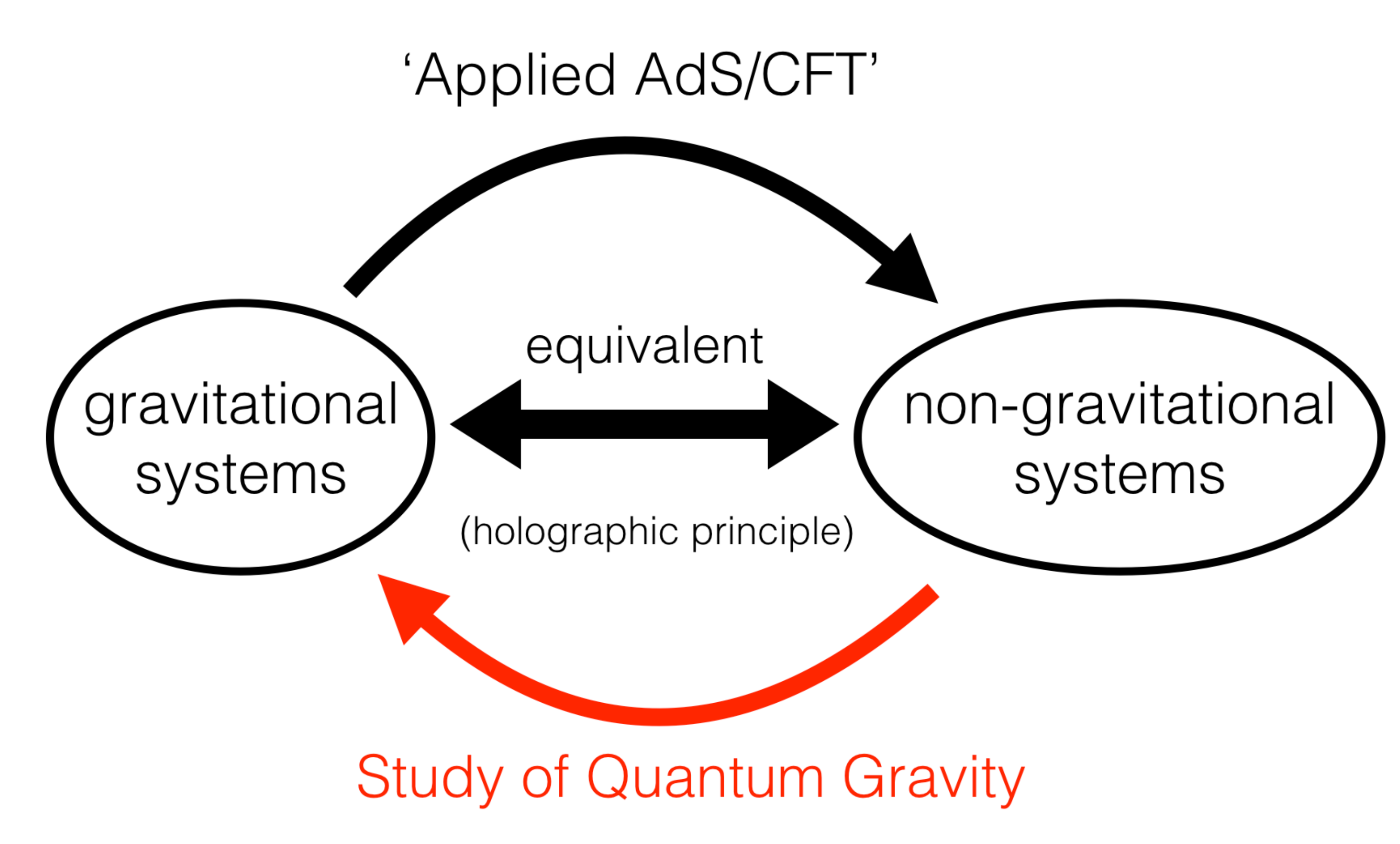}}}
\caption{Via the holographic principle, quantum gravitational systems can be defined by using 
non-gravitational systems. By studying the latter, for example by the lattice simulation, 
we can learn about the former. 
 }
\label{duality}
\end{center}
\end{figure}

The key concept from superstring theory, which the majority of the audience may not be familiar with, 
is {\it holographic principle}.
Superstring theory has been studied as a promising candidate of the theory of quantum gravity. 
One of the biggest motivations to study quantum gravity is the evaporation of the black hole due to the quantum effects \cite{Hawking:1974rv,Hawking:1974sw}.  
The holographic principle \cite{'tHooft:1993gx,Susskind:1994vu} 
claims the equivalence between black holes, and more general quantum gravitational theories, 
and non-gravitational theories in different spacetime dimensions. 
The gauge/gravity duality conjecture \cite{Maldacena:1997re} 
provides us with explicit examples. 
For example, maximally supersymmetric matrix quantum mechanics (also known as the Matrix Model of M-theory \cite{Banks:1996vh,deWit:1988ig}), 
which is believed to describe a black hole in type IIA superstring theory 
near the 't Hooft large-$N$ limit \cite{Itzhaki:1998dd}, has been studied by using Monte Carlo methods starting in \cite{Anagnostopoulos:2007fw}, and steady progresses have been reported in LATTICE conferences 
\cite{Kadoh:2016eju}. 
The latest results presented by E.~Berkowitz in this conference \cite{Berkowitz:2016ftr} strongly support
the correctness of the duality. 
A generalization with flavors has also been studied recently \cite{Asano:2016xsf}. 
(See also \cite{Forini:2017mpu} for another approach to superstring theory with a lattice method.)
Note that this is not just `analogy' nor `model'; quantum field theories should be literally equivalent to quantum gravitational systems, 
which means they are not distinguishable even in principle. 
More practically, we can regard the non-gravitational systems, which can be formulated nonperturbatively 
by using lattice, as the nonperturbative formulations of the quantum gravitational systems. 
By studying the non-gravitational systems, we can learn about quantum gravity (Fig.~\ref{duality}). 
This is where the lattice meets quantum gravity. 
For a recent review on this subject aimed for lattice practitioners, see \cite{Hanada:2016jok}. 

The other key concept is the {\it quantum simulator on optical lattice} (see e.g. \cite{quantum_simulator_review}).  
Quantum simulators have been proposed in order to solve a class of theories 
to which standard numerical techniques are not feasible. 
In this approach, the model Hamiltonians are physically realized, by trapping the cold atoms 
and turning on appropriate interactions by using the lasers and magnetic field. 
The simulation is an actual experiment. 
%
The quantum simulator approach can be useful for QCD and other quantum field theories,
in order to study the real-time dynamics, 
and in order to introduce the baryon chemical potential without being hampered by the sign problem; 
see e.g. \cite{Wiese:2014rla,Zohar:2015hwa}. 

\begin{figure}[htbp]
\begin{center}
\rotatebox{0}{
\scalebox{0.8}{ 
\includegraphics[width=10cm]{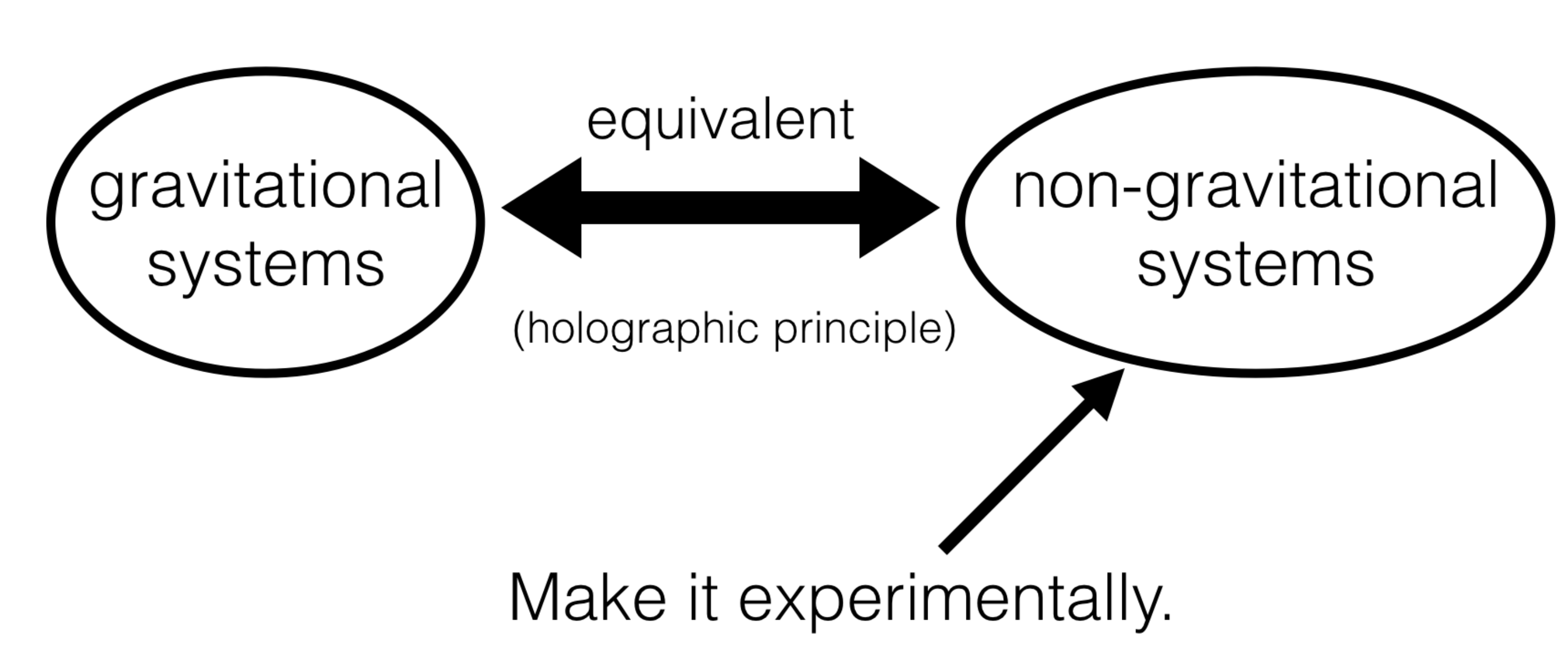}}}
\caption{The realization of the non-gravitational systems which are dual to the gravitational systems
via the holography can lead to the experimental study of quantum gravity. 
 }
\label{EQG}
\end{center}
\end{figure}

The combination of these concepts naturally leads to the idea of {\it the experimental quantum gravity} --- 
if field theory duals of quantum gravitational systems can be realized experimentally, 
that is equivalent to the experimental realization of quantum gravitational systems (Fig.~\ref{EQG}). 
Of course such quantum field theories are very complicated and it is not easy to actually engineer them
at this moment. Recently, however, a very good candidate has been proposed: 
the Sachdev-Ye-Kitaev (SYK) model \cite{Sachdev:2015efa,Kitaev_talk}. 
The SYK model is a simple quantum mechanical system consisting of $N$ fermions, 
which has been proposed as a dual of a black hole in two-dimensional anti de-Sitter space. 
In this presentation, we show a recipe to realize the SYK model on an optical lattice \cite{Danshita:2016xbo}.
For a related proposal, see \cite{Garcia-Alvarez:2016wem}. 

\section{Sachdev-Ye-Kitaev model}
The Sachdev-Ye-Kitaev (SYK) model \cite{Sachdev:2015efa,Kitaev_talk} is a model of $N$ spinless fermions. 
The Hamiltonian is\footnote{
This Hamiltonian is the one with complex fermion \cite{Sachdev:2015efa}. 
Another one with real fermions \cite{Kitaev_talk} is often studied as well.  
They are equivalent in the large-$N$ limit with fixed $J$, after the disorder average. 
} 
\begin{eqnarray}
H=\frac{1}{(2N)^{3/2}}\sum_{ijkl}J_{ij;kl}\hat{c}^\dagger_i\hat{c}^\dagger_j\hat{c}_k\hat{c}_l, \label{Hamiltonian_SYK}
\end{eqnarray}
where indices $i,j,k,l$ run from $1$ to $N$. 
The creation and annihilation operators $\hat{c}^\dagger_i$ and $\hat{c}_i$ satisfy the usual anti-commutation relations 
\begin{eqnarray}
\{\hat{c}_i,\hat{c}_j\}=\{\hat{c}^\dagger_i,\hat{c}^\dagger_j\}=0, 
\qquad
\{\hat{c}^\dagger_i,\hat{c}_j\}=\delta_{ij}.  
\end{eqnarray}
$J_{ij;kl}$ is a complex random coupling constant which satisfies
\begin{eqnarray}
J_{ij;kl}
=
-J_{ji;kl}
=
-J_{ij;lk}, 
\end{eqnarray}
\begin{eqnarray}
J_{ij;kl}
=
J_{kl,ij}^\ast
\end{eqnarray}
and
\begin{eqnarray}\
\overline{({\rm Re}~J_{ij;kl})^2}=\begin{cases}J^2/2&(\{i,j\}\neq\{k,l\})\\J^2&(\{i,j\}=\{k,l\})\end{cases},\hspace{2em}
\overline{({\rm Im}~J_{ij;kl})^2}=\begin{cases}J^2/2&(\{i,j\}\neq\{k,l\})\\0&(\{i,j\}=\{k,l\})\end{cases}.
\end{eqnarray}
The bar $\overline{\ \cdot\ }$ stands for the disorder average.  
The dimensionless combination of the coupling $J$ and temperature $T$, 
which is $J/T$, is a good expansion parameter, and the strong coupling and low temperature are the same.   
In the following, we take $J$ as the unit of energy.

At large-$N$ and strong coupling, the SYK model has striking properties resembling a black hole. 
Firstly, Sachdev \cite{Sachdev:2015efa} pointed out the agreements of the entropy and several correlation functions. 
Furthermore, Kitaev \cite{Kitaev_talk} has shown that the Lyapunov exponent satisfies 
the pattern proposed by Maldacena, Shenker and Stanford \cite{Maldacena:2015waa} 
for quantum theories with dual gravity description. Namely, the Lyapunov exponent 
takes the value $2\pi T$ at strong coupling limit $J/T\to\infty$, 
while at weaker coupling it is smaller than $2\pi T$. 
For further progresses, see e.g. \cite{Maldacena:2016hyu,Maldacena:2016upp,Jensen:2016pah}.

In order to simplify the experimental realization, we slightly modify the SYK model, 
by changing the complex random coupling $J_{ij,kl}$ to be real: 
\begin{eqnarray}
J_{ij;kl}
=
-J_{ji;kl}
=
-J_{ij;lk}, 
\end{eqnarray}
\begin{eqnarray}
J_{ij;kl}
=
J_{kl,ij}. 
\end{eqnarray}
The new disorder average is
\begin{eqnarray}
\overline{|J_{ij;kl}|^2}
=\begin{cases}J^2&(\{i,j\}\neq\{k,l\})\\2J^2&(\{i,j\}=\{k,l\})\end{cases},
\end{eqnarray}
 and for $\{i,j\}\neq\{k,l\}$
\begin{eqnarray} 
\overline{J_{ij,kl}J_{pq,rs}}=J^2\left\{
\left(\delta_{ir}\delta_{js}-\delta_{is}\delta_{jr}\right)
\left(\delta_{kp}\delta_{lq}-\delta_{kq}\delta_{lp}\right)
+
\left(\delta_{ip}\delta_{jq}-\delta_{iq}\delta_{jp}\right)
\left(\delta_{kr}\delta_{ls}-\delta_{ks}\delta_{lr}\right)
\right\}.  
\label{eq:realJJave}
\end{eqnarray} 
The normalization has been chosen so that the energy distribution at large-$N$ coincides with that of the original SYK model.
The second term inside $\{\cdots\}$ in \eqref{eq:realJJave} is absent in the original SYK model. Due to this, each Feynman diagram contains additional terms after disorder average.
However they are $1/N$-suppressed in general, 
and hence this modified model agrees with the original model at large-$N$.     
In the following, we will consider this `real' SYK model. 

\section{Toward experimental quantum gravity}

One of the hardest steps for realizing the SYK model in optical-lattice experiments is the implementation of the all-to-all two-body hopping, because the interactions on lattice systems
are intrinsically `local'
in the sense that the motion of the particles is dominated by the nearest-neighboring one-body hopping. 
In order to overcome this difficulty, we consider a coupling of two atoms with $n_s$ molecular states, described by the following Hamiltonian with real Gaussian random couplings $g_{s,ij}$:
\begin{eqnarray}
\hat{H}_{\rm m}
=
\sum_{s=1}^{n_s}\left\{\nu_s \hat{m}_s^\dagger\hat{m}_s
+
\sum_{i,j}g_{s,ij}\left(
\hat{m}_s^\dagger\hat{c}_i\hat{c}_j
-
\hat{m}_s\hat{c}_i^\dagger\hat{c}_j^\dagger
\right)\right\}.  
\label{Hamiltonian_multi_molecular_states}
\end{eqnarray}
By integrating $\hat{m}_s$ out, we obtain the following effective Hamiltonian,
\begin{eqnarray}
\hat{H}_{\rm eff}
=
\sum_{s,i,j,k,l}\frac{g_{s,ij}g_{s,kl}}{\nu_s}\hat{c}^\dagger_i\hat{c}^\dagger_j\hat{c}_k\hat{c}_l. 
\label{Hamiltonian_AMO}
\end{eqnarray}
(A similar model has been considered in \cite{Bi:2017yvx}.)
Previously, a similar way of designing a kind of two-body hopping term by means of intermediate two-particle states has been proposed in \cite{Buchler:2005fq}. 


Let us take $\nu_1=\nu_2=\cdots=\nu_{n_s}\propto \sqrt{n_s}$. 
When $n_s$ is large enough, $\sum_{s}\frac{g_{s,ij}g_{s,kl}}{\nu_s}$ should become Gaussian except for the diagonal elements $(i,j)=(k,l)$ or $(i,j)=(l,k)$ (note that $g_{s,ij}^2$ is always positive), 
because it is simply an $n_s$-step random walk for each set of indices $(i,j,k,l)$.
In order to improve the behavior of the diagonal elements, we take $n_s$ to be even, 
and set $\nu_s = +\sqrt{n_s} \sigma_\mathrm{s}$ for even $s$ and $\nu_s = -\sqrt{n_s} \sigma_\mathrm{s}$ for odd $s$. 
We take the variance of $g_{s,ij}$ to be $\sigma^2 = \sigma_\mathrm{g}^2$,
with $\sigma_\mathrm{g}^2 / \sigma_\mathrm{s} = J/(2N)^{3/2}$.
In the following we set $\sigma_\mathrm{s} =\sigma_{\rm g}= J/(2N)^{3/2}$.
It is not hard to see that
the properties needed in the real-SYK model are satisfied at $n_s=\infty$. 
We identify $\sum_{s}\frac{g_{s,ij}g_{s,kl}}{\nu_s}$
defined in this way with $J_{ij,kl}/(2N)^{3/2}$.

\begin{figure}[htbp]
\begin{center}
\rotatebox{0}{
\scalebox{0.8}{ 
\includegraphics[width=10cm]{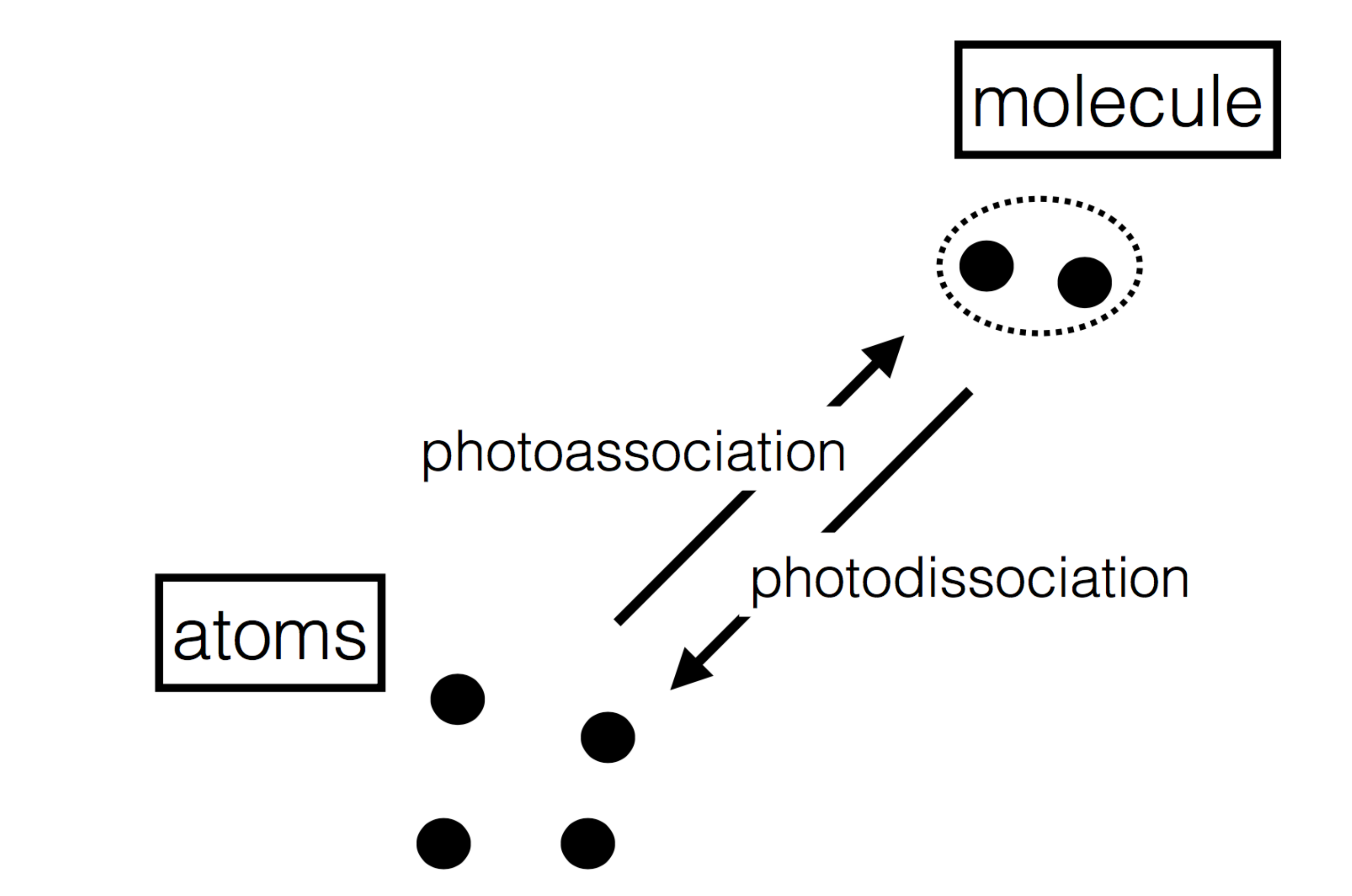}}}
\caption{
By introducing laser beams with appropriate frequencies, 
the photoassociation and photodissociation processes, 
which can be regarded as the interaction between a molecule and two atoms 
in \eqref{Hamiltonian_multi_molecular_states}, can be introduced. 
 }
\label{PA}
\end{center}
\end{figure}

The model (\ref{Hamiltonian_AMO}), which is equivalent to the SYK model at $n_s=\infty$, 
can in principle be created in a system of optical lattices loaded with ultracold gases \cite{Danshita:2016xbo}. 
In the proposed scheme, we utilize the photoassociation and photodissociation processes that coherently convert two atoms into a bosonic molecule in a certain electronic (or hyperfine), vibrational, and rotational state~\cite{jones-06}, and vice versa (Fig.~\ref{PA}). For details, see \cite{Danshita:2016xbo}. 

\section{Discussions and future directions}
In this presentation, we have suggested a path to create a quantum black hole in the laboratory, 
by experimentally realizing the SYK model. Of course, that we can make a black hole {\it in principle} 
does not mean we can make it immediately. 
There are many technical challenges in actual experiments, as summarized in \cite{Danshita:2016xbo}. 
Also, from the theoretical side, it is desirable to simplify the theory. 
One interesting aspect of the SYK model is that various different theories lead to the same 
large-$N$ limit. In this presentation, a few examples have been explained. 
Historically, Sachdev considered another model (Sachdev-Ye model) first, 
and then it was realized that the SYK model gives the same results. 
Yet another equivalent model without disorder has been found in \cite{Witten:2016iux}. 
Hopefully, there is a nice form which allows easier experimental implementation. 
It is also interesting to pursue other theories with dual gravitational interpretations. 
\section*{Acknowledgments}
The authors thank S.~Aoki, G.~Gur-Ari, S.~Nakajima, M.~Sheleier-Smith, S.~Shenker, B.~Swingle, and Y.~Takahashi for discussions. Discussions during the YITP workshop (YITP-W-16-01) on ``Quantum Information in String Theory and Many-body Systems" were useful to complete the paper \cite{Danshita:2016xbo}. The authors acknowledge KAKENHI grants from JSPS: Grants No.~JP15H05855 (I.D. and M.T.), No.~25220711 (I.D.), No.~25287046 (M.H.), and No.~26870284 (M.T.). I.D. was supported by research grant from CREST, JST.
In addition, 
M.~H. thanks B.~Lucini for his help during the editorial process. 
Although in the end it did not work out because of the delay in the publication of \cite{Danshita:2016xbo}, his help was certainly useful. 

\end{document}